# Local nature of 0.1 Hz oscillations in microcirculation is confirmed by imaging photoplethysmography

Irina A. Mizeva[1*], Natalia P. Podolyan[2], Oleg V. Mamontov[3,4], Anastasiia V. Sakovskaia[5], and Alexei A. Kamshilin[2]

[1]Institute of Continuous Media Mechanics of Ural Branch of the Russian Academy of Sciences, Perm, Russia.

[2]Laboratory of New Functional Materials for Photonics, Institute of Automation and Control Processes of Far East Branch of the Russian Academy of Sciences, Vladivostok, Russia.

[3]Department of Circulation Physiology, Almazov National Medical Research Centre, Saint Petersburg, Russia.

[4]Department of Departmental Therapy, Pavlov First Saint Petersburg State Medical University, St. Petersburg, Russia

[5]Institute of Therapy and Instrumental Diagnostics, Pacific State Medical University, Vladivostok, Russia.

*Corresponding Author: Irina A. Mizeva, e-mail: mizeva@icmm.ru,

614013, Ac. Koroleva 1, Perm, Russia

Institute of Continuous Media Mechanics UrB RAS

E-mail addresses: mizeva@icmm.ru (IAM), podolian@iacp.dvo.ru (NPP), mamontoffoleg@gmail.com (OVM), sakovska86@mail.ru (AVS), alexei.kamshilin@yandex.ru (AAK)

**Abstract**

Low-frequency oscillations in the human circulatory system is important for basic physiology and practical applications in clinical medicine. Our objective was to study which mechanism (central or local) is responsible for changes in blood flow fluctuations at around 0.1 Hz. We used the method of imaging photoplethysmography synchronized with electrocardiography to measure blood-flow response to local forearm heating of 18 healthy male volunteers. The dynamics of peripheral perfusion was revealed by a correlation processing of photoplethysmography data, and the central hemodynamics was assessed from the electrocardiogram. Wavelet analysis was used to estimate the dynamics of spectral components. Our results show that skin heating leads to multiple increase in local perfusion accompanied by drop in blood flow oscillations at 0.1 Hz, whereas no changes in heart rate variability was observed. After switching off the heating, perfusion remains at the high level, regardless decrease in skin temperature. The 0.1 Hz oscillations are smoothly recovered to the base level. In conclusion, we confirm the local nature of fluctuations in peripheral blood flow in the frequency band of about 0.1 Hz. A significant, but time-delayed, recovery of fluctuation energy in this frequency range after cessation of the skin warming was discovered. This study reveals a novel factor involved in the regulation microcirculatory vascular tone. A comprehensive study of hemodynamics using the new technique of imaging photoplethysmography synchronized with electrocardiography is a prerequisite for development of a valuable diagnostic tool.

# 1. Introduction

The human cardiovascular system is characterized by the presence of oscillations in a wide range of frequencies. The central and peripheral circulation are under the control of a multilevel system, including neurohumoral regulation by the autonomic nervous system. Regulation of blood circulation due to variations in vascular tone provides an adaptive response of the cardiovascular system to physiologic and pathologic factors [1]. At the functional level of vascular tone regulation, spectral components with a frequency lower than respiratory ($<0.2$ Hz) are of particular interest [2]. The crucial importance of blood flow fluctuations around 0.1 Hz has been repeatedly emphasized in works related to the brain physiology as entrainment of vasomotion links neuronal pathways to functional connections [3], and plays the key role in the purification of brain tissue from metabolites in the glymphatic system of the brain [4]. Therefore, the study of low-frequency oscillations in the human circulatory system is of great importance for fundamental physiology, practical application in clinical medicine.

The origin of oscillations at a frequency close to 0.1 Hz in cardiovascular system remains uncertain and involve different physiological processes depending on blood vessel morphology. On the one hand, the presence of such oscillations in the spectrum of human heart rate (HR) is established. HR fluctuations with the frequency of 0.1 Hz, associated with spontaneous variations of arterial pressure, are called Mayer waves [2]. Closely related to synchronous oscillations of efferent sympathetic nervous activity, Mayer waves are almost invariably enhanced during sympathetic activation. Some early studies have suggested that these waves result from the activity of an endogenous oscillator located either in the brainstem or in the spinal cord [5]. More recent studies, based on the effects of sinoaortic baroreceptor denervation, dispute this point of view [6,7]. Several theoretical models of dynamic arterial pressure control have been developed to predict Mayer waves [8–11]. In these models, it was assumed that the multiple dynamic components and time delays present in the baroreflex loop would produce a resonant, self-supported arterial pressure oscillation. Analysis of the various transfer functions of the rat baroreceptor reflex reveals that Mayer waves are transient oscillatory responses to hemodynamic perturbations rather than true feedback oscillations [12]. However, at the moment, the issue of Mayer waves source is not definitely understood [2].

On the other hand, spontaneous rhythmic fluctuations in the blood vessel diameter were observed at the same frequency of 0.1 Hz, which are called vasomotions and are presumably generated locally without the need for external drivers [13]. Being for the first time detected in bat wing veins [14], vasomotions has been described in many other vessels [15]. Primarily vasomotions occur in small arteries and arterioles [16–18] and has been observed both *in vivo* and *in vitro* [15,19]. The frequency and amplitude of these oscillations are related to the vessel size [13]. The frequency of vasomotions observed *in vitro* is consistent with that *in vivo* in the myogenic frequency range from 0.05 to 0.14 Гц [20].

To investigate vasomotions in peripheral blood vessels and associated reaction of the cardiovascular system, it is necessary to use local testing, such as a test for local heating of the skin to ≈42°C [21]. In this test, measurements of blood flow parameters are usually carried out by an optical method, namely the laser Doppler flowmetry (LDF) [21,22]. However, the sensors in this method usually come into contact with the skin, which affects hemodynamic processes thus changing the measured parameters of blood flow [23]. Moreover, blood flow in LDF is characterized by relative perfusion units, the physical meaning of which has a controversial interpretation [24].

Imaging photoplethysmography (iPPG) is a novel contactless method for measuring tissue perfusion that have drawn attention of researchers due to the capability to assess at once blood flow parameters over large area [25–27]. The latter advantage is especially useful since it allows one to assess spatial distribution of blood-flow parameters and its dynamics, which may vary with long-term measurements. A modified technique consisting of synchronized registration of iPPG and electrocardiogram (ECG) was recently proposed in our group [28]. More recently we applied this method to monitor the perfusion response to local heating of the forearm while controlling all factors that may affect microcirculation [29]. It was shown that the local heating of the skin results in a multiple increase of the perfusion, and the reaction to local hyperemia has a well-known biphasic dynamics previously identified by the LDF method [21,30,31]. During the local heating test, there are significant changes in vascular tone due to complete relaxation of microvascular smooth muscle cells, leading to the achievement of a local maximum perfusion [22]. Consequently, the reaction of the muscle component to local heating can serve as a marker of the state of peripheral blood vessels [32]. Continuous long-term assessment of perfusion is a prerequisite for spectral analysis, which allows one to reveal several bands of low-frequency oscillations in blood flow and analyze their dynamics during physiological tests [33]. Among spectral methods, wavelet analysis has proven itself to be reliable for processing peripheral blood flow data. Blood flow measurements by LDF revealed several low-frequency bands corresponding to oscillations in vascular tone attributed to endothelial (0.01 - 0.02 Hz), neurogenic (0.02 - 0.05 Hz) and myogenic (0.05 - 0.14 Hz) mechanisms of vascular tone regulation [33,34]. In our recent study of local hyperemia assessment by the modified iPPG technique, a wavelet analysis was applied to identify the difference in endothelial fluctuations between smokers and non-smokers [32].

Since the modified iPPG technique has a distinctive feature of simultaneous recording of ECG (determined by systemic cardiovascular processes) and video (capable of assessing peripheral vascular tone) [28], the aim of this study was to find out which mechanism (central or local) is responsible for changing in low-frequency fluctuations of blood flow in response to local skin heating.

## 2. Materials and Methods

### 2.1 Participants and study protocol

The study involved 18 healthy male volunteers, whose average age was 45.0 ± 6.2 years. Exclusion criteria were: age less than 18 years, arterial hypertension, smoking, diabetes, cancer, systemic diseases including cardiovascular, respiratory, hepatic and renal failure, metabolic factors, skin diseases and patients with a history of drug dependence or continuous alcohol consumption, which may adversely affect patient compliance with study procedures and outcome. The characteristics of the subjects are given in *Table 1*.

Table 1. Characteristics of subjects involved in the study

| Parameter | Value |
|---|---|
| Number | ($n = 18$) |
| Age (years) | $45.0 \pm 6.2$ |
| Body mass index (kg/m$^2$) | $27 \pm 3$ |
| Systolic blood pressure (mmHg) | $125 \pm 9$ |
| Diastolic blood pressure (mmHg) | $83 \pm 8$ |

Data are reported as mean ± standard deviation.

The site of the study was the Private Healthcare Institution "Central Clinical Hospital Russian Railways Medicine in Vladivostok" and the Institute of Automatics and Control Processes of the Far East Branch of the Russian Academy of Sciences. The study was performed in accordance with the ethical standards presented in the 2013 Declaration of Helsinki, protocol approved by the Interdisciplinary Ethics Committee of Pacific State Medical University, Vladivostok, Russia. All subjects were acquainted with the testing process and signed a written informed consent form.

Measurements were carried out under appropriate sanitary conditions in a darkened room at an air temperature of $23 \pm 1$ °C. The period of subject's adaptation to the conditions of the study was at least fifteen minutes. The subject was comfortably positioned in a special chair with head and leg support to ensure a relaxed posture throughout the experiment. The subject conveniently placed his right hand on the table so that the forearm was at the level of the heart. A transparent heating module was placed on top of the lower third of the forearm (*Figure 1A*). At least two hours before the experiment, subjects abstained from taking food and liquid, as well as substances affecting vasomotor properties of blood vessels. During the experiment, neither talking nor sleeping was allowed.

In this study, we performed an in-depth analysis of experimental data obtained recently for the control group when assessing the reaction of blood flow to local heating [32]. Measurements of each subject lasted from 40 to 60 minutes, during which the video of the heated forearm area was recorded continuously and synchronously with ECG and skin temperature. The study protocol consisted of (i) 5 min baseline recording; (ii) heating to 40 – 42 °C with the rate of about 3.5 °C per minute; (iii) maintaining the maximum skin temperature for 20 minutes; and (iv) switching off the heating while continuing to monitor blood flow and heart rate during relaxation. The recorded video, ECG and skin temperature were uploaded to a personal computer for further offline processing.

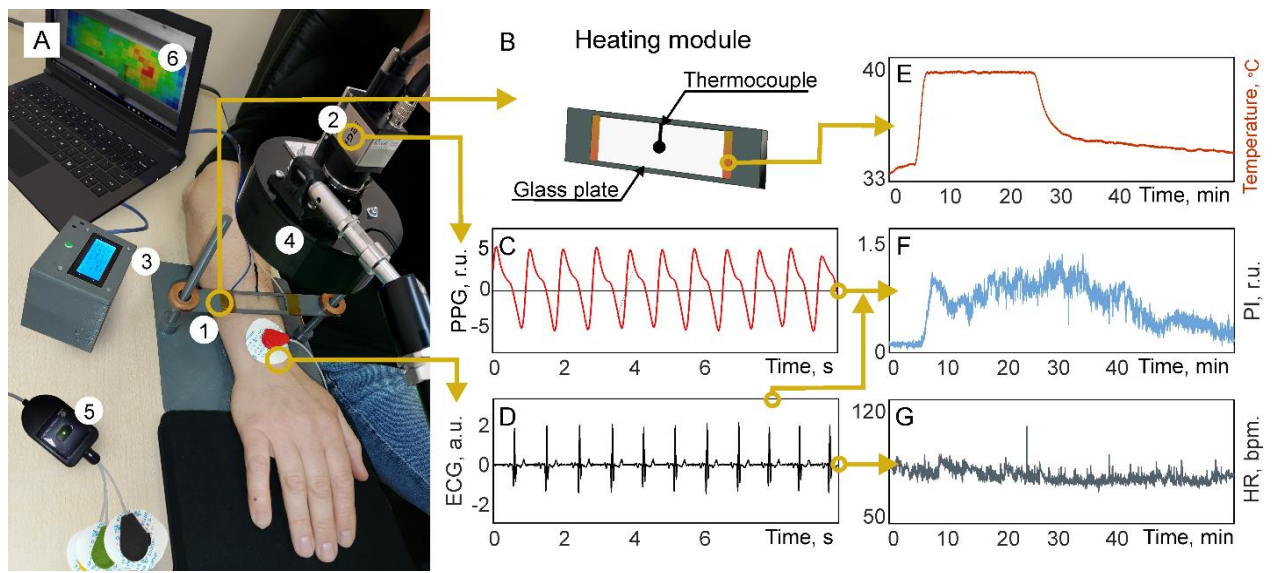

Figure 1. Laboratory system for evaluating perfusion response to local heating. (*A*) Photograph of the imaging photoplethysmography (iPPG) measuring system, which includes a transparent heating module (1), digital monochrome camera (2), skin temperature control unit (3), an illumination module (4), digital electrocardiograph (5), and personal computer (6) for system management and data acquisition and processing. (*B*) Drawing of the heating module. (*C*) An example of the photoplethysmographic (PPG) waveform recorded during 11 cardiac cycles. (*D*) ECG recorded synchronously with the PPG waveform. (*E*) Evolution of the skin temperature during the local heating test. (*F*) Changes in the mean perfusion index during the local heating test. (*G*) Heart rate variability during the local heating test.

## 2.2. Measuring system

To assess the blood flow reaction on local heating, we used a homemade iPPG system [29,32]. A photograph of the experimental setup and typical graphs representing the dynamics of the main variables are shown in *Figure 1*. In our system, the skin was heated using a transparent heating module (3 in *Figure 1A*) and shown in detail in *Figure 1B*. This arrangement allowed us to minimize and stabilize the contact force affecting the cutaneous blood flow due to a sufficiently large contact area of the skin with a 70×20×2 mm$^3$ glass plate. The contact pressure did not exceed 15 mmHg. Video recording of this area by a digital camera (2 in *Figure 1A*) with subsequent data processing makes it possible to compensate for the heterogeneity of the blood flow response to heating. The skin temperature was regulated by a control unit with a feedback system ( 3 ) in which the temperature was measured by a K-type thermocouple placed between the glass and the subject's skin. For better thermal conductivity, Vaseline oil was added to the contact area. The video was recorded at a distance of 20 cm with a monochrome digital camera at a frequency of 36 fps and resolution of 380×256 pixels. The camera was rigidly mounted in the center of an illumination module (4) consisting of four rings of Light-Emitting-Diode's ribbon (250 LEDs operating at the wavelength of 530 ± 25 nm) and mounted on a variable friction support for precise positioning. To improve the signal-to-noise ratio, object illumination and video data recording were performed through a linear-polarizing film with mutually orthogonal orientation of polarization directions for the light source and video camera. Simultaneously, a digital electrocardiograph (5) recorded the patient's ECG in two leads at a frequency of 1 kHz. The recorded video frames and ECG were synchronized in time with the accuracy of 1 ms. Data collection and the management of the entire system were carried out by a personal computer ( 6).

## 2.3. Data processing

Evaluation of perfusion changes during the local heating test was carried out offline using the previously described algorithm [29]. There are four stages of data processing. At the first stage, the recorded images were stabilized using the optical flow algorithm in order to reduce the impact of unavoidable motion artifacts. Already at this stage we exploited the synchronicity of ECG and iPPG recording, which increased the efficiency of image stabilization [28]. The stabilized images were partitioned into Regions of Interest (ROI) of 15×15 pixels that is about 2 by 2 mm at the forearm plane.

At the second stage, we calculated frame-by-frame evolution of the mean pixel value in every ROI over 15 consecutive cardiac cycles every minute. This provides possibility to assess the spatial distribution of the blood pulsation amplitude over the heating area. In iPPG systems, the pulsation amplitude is estimated as the ratio of the PPG-waveform component modulated at the heart rate to the mean value of the waveform, which takes into account possible nonuniformity of the illumination intensity [28,35]. From the resulting spatial distribution, half of the ROIs with a higher pulsation amplitude are selected thus accounting for the heterogeneity of the perfusion distribution. Then, the frame-by-frame evolution of the pixel values averaged over the selected ROIs is calculated and normalized in the similar way, resulting in a PPG waveform, which fragment is shown in *Figure 1C*. This waveform was evaluated during the entire time of the local heating test and used to assess the perfusion dynamics.

At the third stage, we calculated the perfusion index (*PI*) in every cardiac cycle as the difference between the maximum and minimum values of the normalized PPG waveform while determining the boundaries of each cycle by R-peaks of the synchronously recorded ECG. Synchronous video signal acquisition with ECG data is illustrated in *Figure 1* in which the time axes on the graphs on panels *C* (PPG waveform) and *D* (ECG) are linked to each other. An example of the perfusion dynamics assessed during the local heating test is shown in *Figure 1F*. As it was demonstrated in [35] the *PI* parameter reflects the tone of blood vessels supplying the capillary network under study, despite the fact that green light interacts only with capillaries due to the shallow depth of its penetration. This is explained in the alternative model of the PPG signal formation, according to which the modulation of the blood volume in the capillary bed at the heart rate occurs due to mechanical compression/decompression of the density of capillaries by adjacent arteries [36].

While iPPG is an instrument for assessing local changes in blood flow, HR variations reflect systemic effects. An example of HR dynamics, which is evaluated as the inverse time difference between adjacent R-peaks is shown in *Figure 1G*.

At the fourth stage, a wavelet analysis of perfusion dynamics, $PI(t)$, and heart rate, $HR(t)$, was carried out [32]. The spectral composition of $PI(t)$ and $HR(t)$ was evaluated in eleven time intervals, the duration of each of which was 4 min. Of the eleven intervals, the first (designated as *A*) was on the baseline, the intervals from the second to the sixth (*B* to *F*) covered the heating phase, and the rest (*G* to *K*) were in the relaxation phase. For every subject we calculated wavelet coefficients for the entire sequence of both $PI(t)$ and $HR(t)$ [32]. The power spectral densities of the perfusion index $M_i^P(\nu)$ and heart rate $M_i^H(\nu)$ were estimated by integrating the wavelet coefficients in each of time interval $i$. Next, we calculated the normalized spectra of the perfusion index, $NS_i^P$, using normalization by the mean value of the $PI(t)$ in the respective time interval as follows

$$NS_i^P(\nu) = M_i^P(\nu)/\langle PI^2(t)\rangle_i. \tag{1}$$

Since the perfusion index *PI* is proportional to the amplitude of blood flow oscillations at the heart rate, such normalization shows the ratio of the oscillation energy of the selected frequency band to the cardiac wave energy in the studied tissue volume. The normalized spectra of the heart rate dynamics, $NS_i^H(\nu)$, were calculated as

$$NS_i^H(\nu) = M_i^H(\nu)/\langle HR^2(t)\rangle_i. \tag{2}$$

Here $\langle HR^2(t)\rangle_i$ is the mean value of the energy $HR(t)$ in the time interval $i$.

## 2.4. Statistical analysis

The study used methods of non-parametric statistics. To test the hypothesis of the reliability of differences in dependent variables, the Wilcoxon test was utilized. A 95% confidence interval was used. The level of significance of all of the statistical indicators is $P < 0.05$. Statistical analysis of the data was also performed in the Mathematica 10.3 (Wolfram research, USA).

# 3. Results

*Figure 2* shows the typical dynamics of the perfusion and heart rate during the local heating test. The heating results in a multiple increase in $PI(t)$ in a healthy subject. Dependence $PI(t)$ has a typical biphasic shape with fast vasodilation, followed by a nadir and then by a long plateau. After switching off the heating element, skin temperature rapidly decreases whereas the perfusion does not follow the temperature, and in some subjects even increasing. It was found that $PI(t)$ decreases at a very low rate: it does not return to the baseline level even after a long relaxation period (more than 30 minutes).

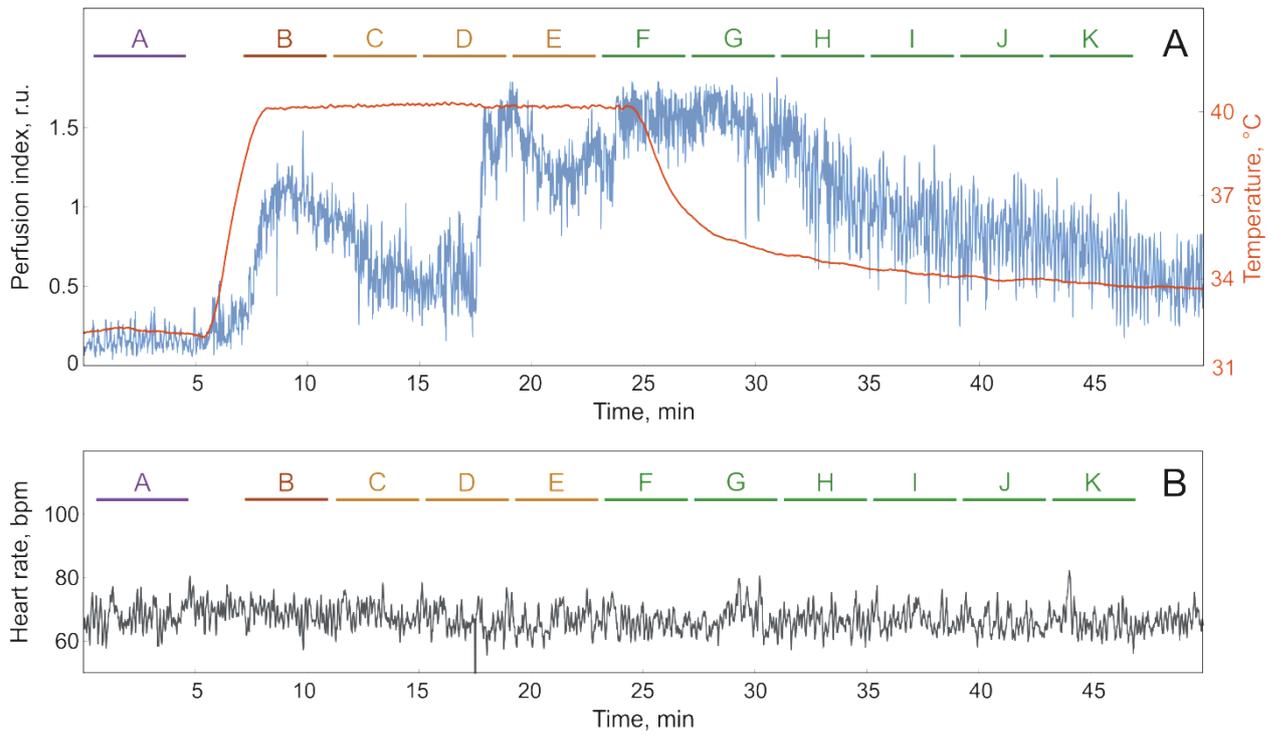

Figure 2. Synchronously recorded dynamics of skin temperature, perfusion, $PI(t)$, and heart rate, $HR(t)$, during a forearm local heating test. (A) An example of simultaneously recorded skin temperature (red curve) and $PI(t)$ (blue curve). (B) Dependence $HR(t)$ synchronously recorded with $PI(t)$. The time intervals in which the wavelet transform of the recorded signals was calculated are shown at the top of each panel.

Comparison of the dynamics of low-frequency components of $PI(t)$ and $HR(t)$ waveforms can be performed using a spectrogram, on which the calculated wavelet coefficients graphically represent the time-frequency waveform sweep. On the spectrogram, the power spectral density components $M(\nu)$ are shown for each moment in time and are color-coded in such a way that red indicates a higher power, while blue corresponds to a lower one.

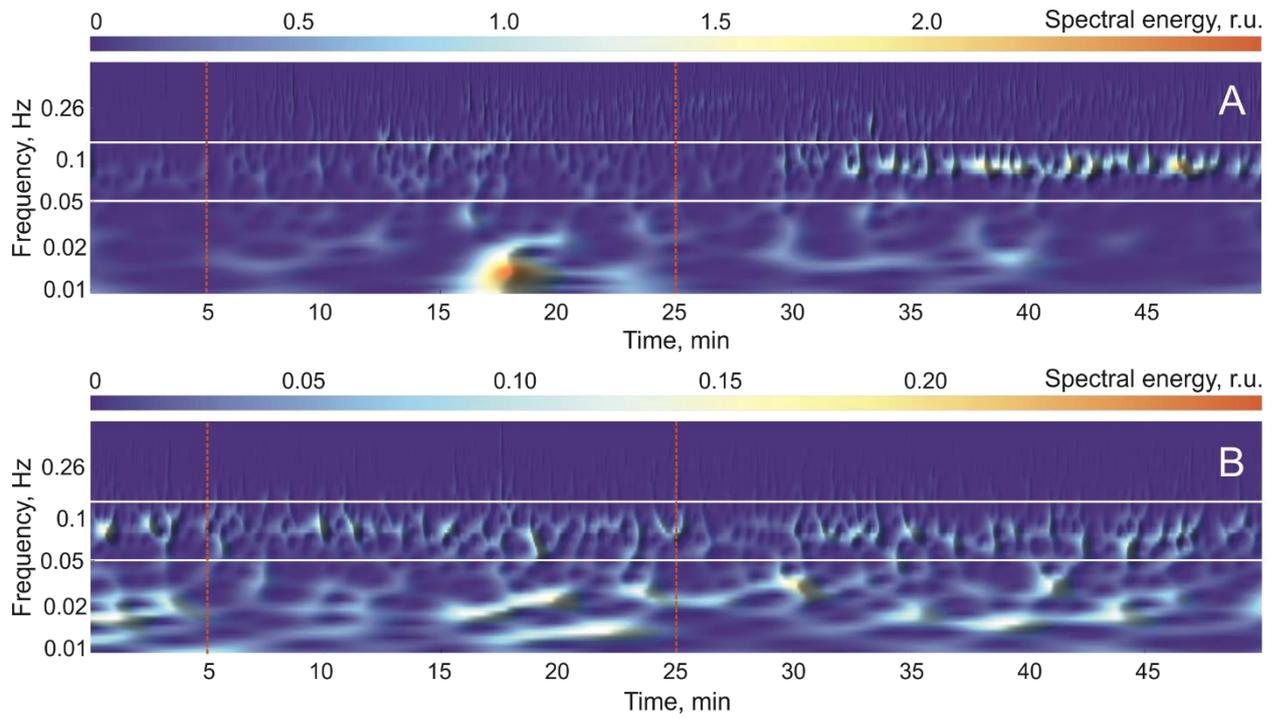

Figure 3. Wavelet spectrograms of the perfusion and heart rate responses to local heating. (*A*) Time-frequency dependence for the perfusion waveform, $PI(t)$. (*B*) Time-frequency dependence for the heart rate, $HR(t)$, Both measured waveforms are shown in *Figure 2*. Red vertical lines mark the moments when the heating module is switched on and off. The horizontal white lines show the frequency range (0.05 - 0.14 Hz).

The wavelet spectrograms (*Figure 3*) of the $PI(t)$ and $HR(t)$ have several notable features. Starting from the 30th minute, an increase in spectral energy of $PI(t)$ in the frequency band of 0.05 - 0.14 Hz is clearly observed (*Figure 3A*), which appears five minutes after turning off the heating element. On the contrary, the spectral energy of $HR(t)$ in this frequency band (as well as at other frequencies) remains unchanged throughout the whole heating test (*Figure 3B*).

To compare the dynamics of the power spectral densities among the subjects, we calculated the normalized spectra $NS_i^P$ and $NS_i^H$ in every time interval (A - K) according to Eqs. 1 and 2, and presented the obtained spectra in *Figure 4* in the form of the box plots for a few representative time intervals. Note that the chosen time intervals correspond to different stages of the heating test. The time interval *A* (0.5-4.5 min), indicated in purple, coincides with the baseline recording; the interval *B* (7-11 min), indicated in brown, is associated with the initial stage of vasodilation, caused by the axon reflex; the intervals *C-F* (11-23 min), marked by orange boxes, correspond to the period during which the heating module remained switched on; and remained stage of the test is covered by the intervals *G-K* (23-55 min), and shown by green boxes. Seeking the clarity of presentation, we plotted in *Figure 4* only four representative intervals *A*, *B*, *E*, and *K* in the frequency range from 0.01 to 0.6 Hz. Perfusion spectra in the selected time intervals are shown in *Figure 4A*. The spectral composition during the baseline (purple boxes) does not statistically differ from that at the end of the test (green boxes). Moreover, a pronounced peak in the band around 0.1 Hz, marked by a vertical blue bar, is clearly seen. In contrast, fluctuations in this frequency band are significantly lower during the heating (brown and orange boxes). It is worth noting that spectral properties in these frequencies are similar during both phases of vasodilation.

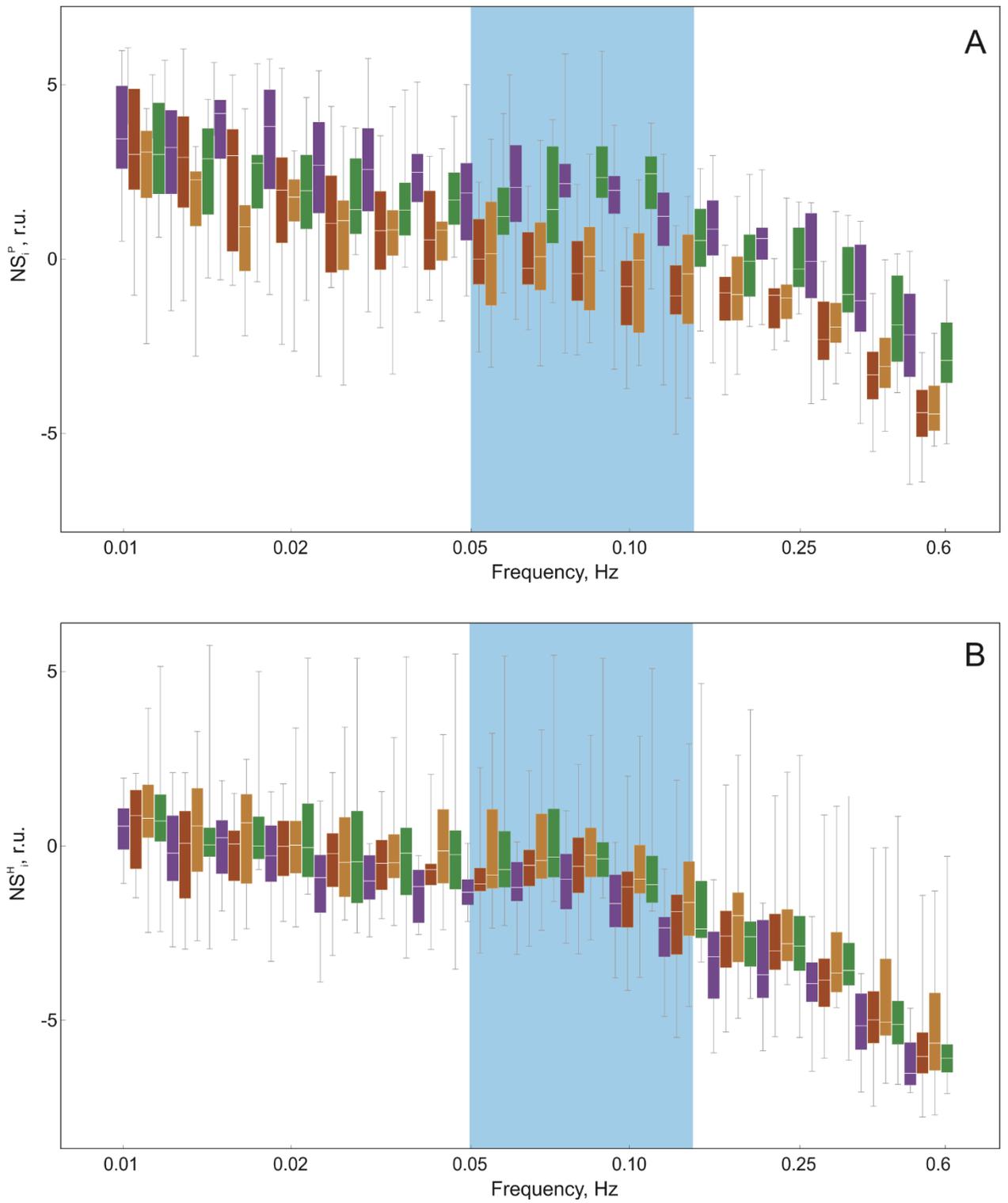

Figure 4. Normalized spectral energy of the perfusion index and heart rate ($NS_i^P$ and $NS_i^H$, respectively) in four representative time intervals, averaged over all subjects under study. (A) Wavelet spectra $NS_i^P$ of the perfusion index, $PI(t)$. (B) Wavelet spectra $NS_i^H$ of the heart rate, $HR(t)$. The representative time intervals in the heating test $A$ (0.5 - 4.5 min), $B$ (7 - 11 min), $E$ (19-23 min), and $K$ (43-47 min) are colored purple, brown, orange, and green, respectively. Blue vertical bars indicate the frequency band of 0.05 – 0.14 Hz. The data are presented in the logarithmic scale.

Unlike $PI(t)$, the wavelet spectra of $HR(t)$, assessed in the same time intervals, does not show any statistically significant difference as one can see in *Figure 4B*. Considering observed peculiarities, we focused our analysis on the frequency band from 0.05 to 0.14 Hz, which is centered to 0.1 Hz.

By averaging the normalized spectra $NS_i^P$ in the frequency range of 0.05-0.14 Hz in each of the time intervals, we obtain 11 values to demonstrate the dynamics of changes in the mean spectral energy of perfusion variations in the frequency band around 0.1 Hz. Similarly, 11 more values were calculated for the normalized heart rate spectra, $NS_i^H$, in the same frequency band. The statistical distribution of the spectral components of $PI(t)$ and $HR(t)$ for the whole group is presented as box-whisker diagram in *Figure 5* by blue and dark-grey boxes, respectively. As one can see in *Figure 5*, the dynamics of the energy of myogenic fluctuations in the *PI(t)* (peripheral hemodynamics) differs significantly from that in the *HR(t)* representing the central hemodynamics.

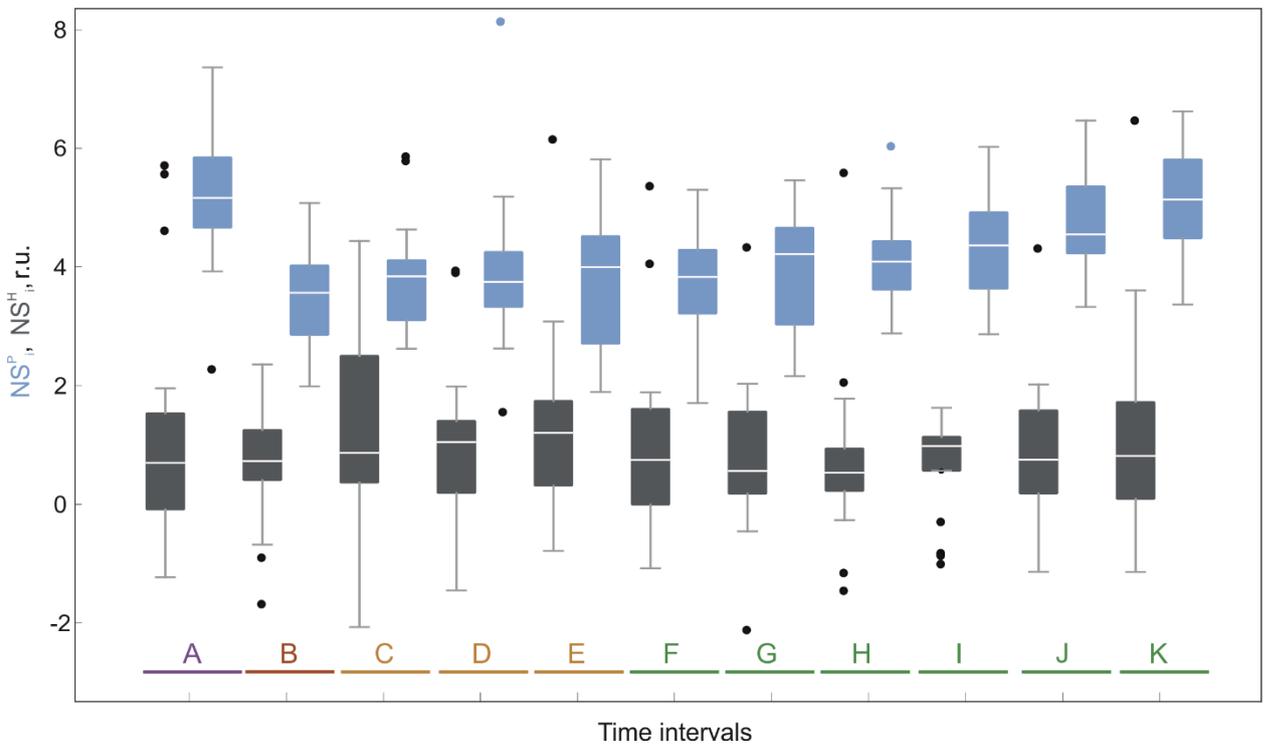

Figure 5. Dynamics of normalized spectral components of PI(t) ($NS_i^P$, shown by blue boxes) and HR(t) ($NS_i^H$, shown by dark-grey boxes) in the frequency band of 0.05-0.14 Hz for all subjects. The spectral components are represented on a logarithmic scale. The position of the time intervals on the time scale of the local heating test is shown in *Figure 2*.

While no statistically significant difference in the myogenic spectral energy of $HR(t)$ between two different time intervals was found in any pair of time intervals, the pattern is completely different for spectral energy of $PI(t)$. *Table 2* complements *Figure 5* and shows the magnitude of the statistically significant difference in the myogenic spectral energy of the $PI(t)$ calculated for all possible pairs of the time intervals. It is seen in *Figure 5* that the energy of perfusion fluctuations in the myogenic frequency band significantly decreases immediately after the start of skin heating (time interval *B* compared to *A*). Between these intervals, there was a fivefold decrease in group median spectral energy with $P = 0.0002$ (*Table 2*).

Table 2. Magnitude of the significance of difference between the median parameter $NS_i^P$ assessed in different pairs of the time intervals.

| Time intervals | Time intervals | | | | | | | | | |
|---|---|---|---|---|---|---|---|---|---|---|
| | B | C | D | E | F | G | H | I | J | K |
| A | 0.0002* | 0.003* | 0.0003* | 0.002* | 0.0009* | 0.003* | 0.0002* | 0.0004* | 0.007* | 0.08 |
| B | | 0.38 | 0.12 | 0.43 | 0.23 | 0.067 | 0.004* | 0.006* | 0.0002* | 0.0002* |
| C | | | 0.51 | 0.76 | 0.76 | 0.43 | 0.12 | 0.023* | 0.001* | 0.0002* |
| D | | | | 0.6 | 0.7 | 0.25 | 0.067 | 0.045* | 0.01* | 0.006* |
| E | | | | | 0.48 | 0.097 | 0.067 | 0.055 | 0.011* | 0.003* |
| F | | | | | | 0.23 | 0.11 | 0.055 | 0.02* | 0.002* |
| G | | | | | | | 0.43 | 0.074 | 0.045* | 0.0009* |
| H | | | | | | | | 0.09 | 0.002* | 0.003* |
| I | | | | | | | | | 0.09 | 0.002* |
| J | | | | | | | | | | 0.97 |

\* - statistically significant difference ($P < 0.05$).

After first drop, the energy of perfusion fluctuations stays on this low level during all time the heater is switched on. The difference between all intervals while the heater switched on is nonsignificant as seen in *Table 2*. After turning off the heater, myogenic oscillations gradually intensify and their significant changes are confirmed by a statistically significant difference in pairs (*G - I, H - J, I - K*). This means that at least over every 8 minutes the spectral energy increases significantly. The difference becomes nonsignificant only in the relaxation time interval of *K* corresponding to 43-47 min or the heating test. Therefore, the oscillation restores only after 30 min of relaxation. This result is proved by statistically equivalence of the intervals *A* and *K*: $P = 0.8$ (*Table 2*).

# 4. Discussion

In this study, we used a combination of two time-synchronized experimental modalities, iPPG and ECG, to assess the response of the cardiovascular system to the local heating test. While the former method allows us to reveal the features of peripheral hemodynamics, the latter shows the contribution of the autonomic nervous system to the centralized regulation of hemodynamics. Our study has confirmed previously reported observations [21,30,31] that mild heating of a forearm skin area leads to a multiple increase in perfusion. Such kind of vasodilation results in almost complete disappearance of rhythmic oscillations in blood cell flux at a frequency of around 0.1 Hz, which was first discovered in 1987 by Kastrup et al. [37]. and confirmed in this study, as well. The wavelet analysis of blood flow dynamics during the local heating test revealed a sharp, statistically significant drop in the energy of blood flow fluctuations in the frequency range from 0.05 to 0.14 Hz. Significant decrease in the energy of blood flow fluctuations in this spectral band in response to local heating was also reported by Geyer et al. [38], although they used a slower skin heating protocol. .

Variations in the tone of small vessels due to contraction and relaxation of muscular components of the vascular wall play a significant role in the regulation of skin blood flow. Under the conditions of performing the test of prolonged local hyperemia in the first minutes of heating, there is an increase in blood flow due to relaxation of vascular smooth muscle and a decrease in sympathetic influence on the vascular wall of arterioles, which, as a consequence, leads to a multiple increase in perfusion (vasodilation due to the axon reflex). The rapid increase in blood flow in the vascular bed creates shear stress on the luminal surface of the vessels. In response to this stimulus, endothelial cells react by increasing the synthesis of primarily nitric oxide, prostacyclin, bradykinin [39]. Increased production of these substances leads to a decrease in the concentration of intracellular $Ca^{2+}$ and relaxation of smooth muscle of the vessel wall [40]. Subsequent heating increases nitric oxide release by causing a mild inflammatory reaction, in which local chemical mediators increase the permeability of capillaries and postcapillary venules, which also leads to vasodilation of resistive vessels. Despite the difference in mechanism of dilatation between the first and second phases of the response to skin heating, we did not find a statistically significant difference in blood flow fluctuations at frequencies of around 0.1 Hz between these phases, (*Figure 5* and *Table 2* for the time-intervals *B – E)*. These fluctuations of the blood flow remain at a very low level for 5 minutes even after turning off the heating element and only then begin to gradually recover to baseline values. It is worth noting that the restoration of fluctuations in this frequency band up to the baseline level takes more than 25 minutes for all subjects.

At the same time, after turning off the heater, we observed continued dilatation, PI did not restore up to the initial level up to the end of our measurements (*Figure 2*). It should be emphasized that there is a high level of prolonged perfusion regardless of a sharp (within 2 minutes) drop in skin temperature. Such a significant discrepancy between the dynamics of skin temperature and evolution of perfusion after switching off the external heating was first reported by Minson et al., who called this observation paradoxical and left it without interpretation [22]. The protocol of the local heating test with the cessation of skin warming is rarely used, but there are few reports of continued vasodilation after a sharp decrease in skin temperature [25,29,32]. Our study has revealed for the first time that in the post-heating period, simultaneously with continued vasodilation, there is a significant increase in the energy of blood flow fluctuations in the low-frequency range of 0.05-0.14 Hz. Such an increase in myogenic fluctuations has not been reported in previous studies [37,38]. We speculate that the gradual increase in these oscillations and prolonged perfusion after turning off the heater are interrelated phenomena.

The spectral analysis of heart rate dynamics, associated with the central mechanism of hemodynamics regulation, did not reveal any changes in the energy of HR fluctuations in this frequency range throughout the test. Therefore, we suggest that the observed activation of blood flow fluctuations at the frequency of about 0.1 Hz after cessation of heat exposure is unlikely to be associated with the regulation of vascular tone by centrally originated mechanisms. Our experimental data confirm the theoretically established fact that a small change in microcirculatory parameters can lead to either suppression or significant enhancement of blood flow fluctuations at the frequency of about 0.1 Hz [41]. As suggested, such behavior of the microcirculatory system is due to the strong nonlinearity of the response of smooth muscle that regulates arteriolar tone to changes in vascular intraluminal pressure [42,43].

# 5. Conclusions

Reliable measurements of blood flow parameters in the forearm area performed by the iPPG system allowed us to experimentally confirm the local nature of fluctuations of peripheral blood flow in the frequency band of around 0.1 Hz. In addition, a significant, but time-delayed, recovery of fluctuation energy in this frequency range after cessation of the skin warming was discovered for the first time.

# 6. Acknowledgments

The authors kindly thank Valeriy V. Zaytsev, Anzhelika V. Belaventseva, and Roman V. Romashko for their valuable help in collecting the experimental data.

## 7. Funding

The work was carried out within the state assignment of IACP FEB RAS, Vladivostok (Theme FWFW-2022-0003) in terms of experiment planning, manufacturing and calibration of a laboratory installation, development of a data processing algorithm and with support of the Ministry of Education and Science of Russia of ICMM UrB RAS, Perm AAAA19-119012290101-5 in terms of the spectral data analysis.

## 8. Author contribution

I.A.M.: Conceptualization, Formal analysis, Software, Writing – original draft. N.P.P.: Conceptualization, Methodology, Formal analysis, Investigation, Data curation, Visualization. O.V.M.: Conceptualization, Methodology, Formal analysis.  A.V.S.: Investigation, Resources. A.A.K.: Conceptualization, Methodology, Software, Data curation, Supervision, Writing – review & editing.

## 9. Conflict of interest

None declared.